\documentclass[letterpaper, 10 pt, conference]{ieeeconf}
\IEEEoverridecommandlockouts
\usepackage{amsmath,amssymb,amsfonts}
\usepackage{blindtext, graphicx}
\PassOptionsToPackage{greek,english}{babel}
\usepackage{textcomp}
\usepackage{xcolor}
\usepackage{booktabs} 
\usepackage{bbding}
\usepackage{soul,color}
\usepackage{comment} 
\usepackage{kantlipsum}
\usepackage{multicol}
\usepackage{microtype}
\usepackage{array}
\usepackage{tabulary}
\usepackage{chngcntr}
\usepackage{multirow}
\usepackage[T1]{fontenc}
\usepackage{amssymb}

\usepackage{tablefootnote}
\usepackage{threeparttable}

\usepackage{cleveref}

\let\Setlength\setlength 
\usepackage{calc}
\newlength{\arrayrulewidthOriginal}

\DeclareMathAlphabet\mathzapf{T1}{pzc}{mb}{it}

\usepackage[style=ieee,minbibnames=1,maxbibnames=2,doi=false,isbn=false,url=false,eprint=false,date=year]{biblatex}
\AtEveryBibitem{\clearfield{pages}}
\AtEveryBibitem{\clearfield{volume}}
\AtEveryBibitem{\clearfield{number}}
\AtEveryBibitem{\clearfield{note}}
\AtEveryBibitem{\clearlist{language}}

\newcolumntype{K}[1]{>{\centering\arraybackslash}p{#1}}
\usepackage{float}
\usepackage[linesnumbered,ruled,vlined]{algorithm2e}
\usepackage{bm}

\newcolumntype{M}[1]{>{\centering\arraybackslash}m{#1}}
\Setlength{\intextsep}{5pt}
\renewcommand{\baselinestretch}{1} 

\IEEEoverridecommandlockouts                              

\begin{document}
\title {Systematic Assessment of Hyperdimensional Computing for Epileptic Seizure Detection \\

\thanks{This work has been partially supported by the ML-Edge Swiss National Science Foundation (NSF) Research project (GA No. 200020182009/1), and the PEDESITE Swiss NSF Sinergia project (GA No. SCRSII5 193813/1).}
\thanks{$^{1}$ U. Pale, T. Teijeiro, and D. Atienza are with the Embedded Systems Laboratory of Swiss Federal Institute of Technology Lausanne, Switzerland.
         {\tt\small \{una.pale, tomas.teijeiro, david.atienza\} @ epfl.ch}}
}


\author{ Una Pale, Tomas Teijeiro, and David Atienza $^{1}$ }

\maketitle

\begin{abstract}
Hyperdimensional computing is a promising novel paradigm for low-power embedded machine learning. It has been applied on different biomedical applications, and particularly on epileptic seizure detection. Unfortunately, due to differences in data preparation, segmentation, encoding strategies, and performance metrics, results are hard to compare, which makes building upon that knowledge difficult. Thus, the main goal of this work is to perform a systematic assessment of the HD computing framework for the detection of epileptic seizures, comparing different feature approaches mapped to HD vectors.
More precisely, we test two previously implemented features as well as several novel approaches with HD computing on epileptic seizure detection. 
We evaluate them in a comparable way, i.e., with the same preprocessing setup, and with the identical performance measures. We use two different datasets in order to assess the generalizability of our conclusions. 
The systematic assessment involved three primary aspects relevant for potential wearable implementations: 1) detection performance, 2) memory requirements, and 3) computational complexity. Our analysis shows a significant difference in detection performance between approaches, but also that the ones with the highest performance might not be ideal for wearable applications due to their high memory or computational requirements. Furthermore, we evaluate a post-processing strategy to adjust the predictions to the dynamics of epileptic seizures, showing that performance is significantly improved in all the approaches and also that after post-processing, differences in performance are much smaller between approaches. 
\end{abstract}


\section{Introduction}
\bstctlcite{IEEEexample:BSTcontrol}

Epilepsy is a chronic neurological disorder that not only impacts a significant portion of the world population (0.6 to 0.8\%)~\cite{mormann_seizure_2007}, by imposing health risks and restrictions on their daily life but also remains an open research question for scientists and doctors. Indeed, one third of the patients still suffer from seizures despite pharmacological treatments~\cite{schmidt_evidence-based_2012}. To provide possible alternatives for such patients, as well as to extract relevant parameters such as frequency, seizure type, or location, EEG (electroencephalography) or IEEG (intracranial EEG) recordings are usually needed. EEG measures the brain's electrical activity on the scalp surface, while IEEG is an invasive technique with surgically placed intracranial electrodes. 

In recent years, with the development of low-power, small-size wearable devices, a range of possible new applications have opened in the medical field in general, as well as for epilepsy patients (e.g. \cite{poh_continuous_2010, beniczky_detection_2013, sopic_e-glass:_2018}). Many of them are based on the idea of continuous 
monitoring with the goal of detecting in real-time or even predicting the upcoming seizures to enable prompt reaction and prevent possible accidents. Such devices have many requirements to satisfy in order to be accepted by patients and doctors: they must be lightweight, small, and unobstructive with significant computational power but also sufficient battery lifetime. This means that many state-of-the-art algorithms
~\cite{emami_seizure_2019, ghosh-dastidar_principal_2008}, are infeasible due to excessive memory and/or power requirements.  

Hyperdimensional (HD) computing is a promising new machine learning (ML) approach inspired by neuroscience~\cite{kanerva_hyperdimensional_2009}, which posits that large but simple circuit networks are fundamental to the brain's computational power. More precisely, neuroscience results suggest that the brain's computation is based on high-dimensional randomized representations of data rather than scalar numerical values~\cite{kanerva_hyperdimensional_2009}. Similarly, HD computing is based on representing data (i.e., feature or channel indexes and values) as vectors with very high dimensionality (usually $>10,000$ values~\cite{kanerva_hyperdimensional_2009}). This enables various properties and operations on such vectors, making it interesting for embedded systems. 

The HD computing framework consists of three stages: encoding, training, and querying. Encoding is the step in which HD vectors are initialized, and data/features are encoded to HD vectors. This step mainly differentiates HD computing approaches among different applications and papers. Training consists of summing up (bundling) all vectors coming from the same class. Then, during the querying step, labels are inferred based on the similarity of the current HD vector with the model vectors representing each class.  

HD computing is a relatively new approach, so only a few but diverse works applied it to biomedical data~\cite{rahimi_hyperdimensional_2016, rahimi_high-dimensional_2017,chang_hyperdimensional_2019, rahimi_hyperdimensional_2020}. 
It has been used for epileptic seizure detection too, by testing different feature approaches~\cite{burrello_one-shot_2018, asgarinejad_detection_2020}. Unfortunately, results are not comparable, as will be discussed in the paper, due to differences in data preparation, segmentation, mapping to HD vectors, and different performance metrics, making it hard to build upon that knowledge. 

Thus, in this work, we contribute to the state of the art in the following manner:
\begin{itemize}
    \item We investigate and systematically compare all features and the methods of encoding them to HD vectors currently available in literature in terms of prediction accuracy (sensitivity, recall, and F1) for both seizure episodes and duration levels, as well as in terms of computational complexity and memory requirements.
    \item We implement new features often used in machine learning techniques for epileptic seizure detection, but which have not yet been implemented in HD computing. 
    \item We utilize two publicly available datasets to improve the generalizability of our methodology.
    \item Among our results, we demonstrate that post-processing significantly reduces the differences between approaches (e.g., from 61.4\% to 7.9\% for episode detection and from 17.4\% to 11.0\% for seizure duration). Moreover, our results show that there are strong trade-offs between approaches in terms of memory footprint (3.8x between largest and smallest footprint) and computational complexity (1056x more operations on HD vectors between most to least computationally complex).
\end{itemize}

\section{Background and Related Work}

HD computing is a novel paradigm interesting for low-power embedded machine learning, due to specific algebraic properties when computing with hyperdimensional vectors: 1) Any randomly chosen pair of vectors is nearly orthogonal, and 2) A vector built from the summation of other vectors is with high probability more similar to its component vectors than any other randomly chosen HD vector. These properties are very useful for classification applications where each class can be represented by a model HD vector, calculated by simply computing an element-wise sum over individual HD vectors of samples belonging to that class. Later, at prediction time, the similarity between an HD vector representing the current sample and model HD vectors for each class are calculated. The sample is then assigned to the class whose model vector has the highest similarity. 
For computationally lightweight applications, binary vectors are used, i.e., each numeric value is 0 or 1, even though the framework supports any numeric value. This enables lower memory requirements and utilizes simpler operations, such as bitwise SUM and XOR operations. 
Hamming distance is used for measuring similarity between vectors (binary or otherwise), along with cosine or dot product distance for integer of floating point vectors.   

The main advantages of HD computing are energy efficiency, lower need for significant data preprocessing, scalability, analysability, no need for expert knowledge for feature and model construction~\cite{rahimi_efficient_2019}, and the potential for one-shot learning~\cite{burrello_one-shot_2018}.
Consequently, different approaches have been proposed and show high potential for biomedical applications. 

Even though HD computing was first proposed in the 1990s, only Kanerva's 2009 overview~\cite{kanerva_hyperdimensional_2009} improved accessibility and interest by illustrating its advantages and potential applications for neuroscience, computer science, mathematics and engineering. Over the following years, the first applications on biosignal processing were proposed~\cite{rahimi_hyperdimensional_2016, rahimi_high-dimensional_2017}. Since then, HD computing has been applied to different challenges: EMG (electromyogram), gesture recognition~\cite{rahimi_hyperdimensional_2016}, emotion recognition from GSR (galvanic-skin response), ECG (electrocardiogram) and EEG~\cite{chang_hyperdimensional_2019}, EEG error related potentials detection~\cite{rahimi_hyperdimensional_2020}, and epileptic seizure detection from iEEG~\cite{burrello_one-shot_2018} and EEG~\cite{asgarinejad_detection_2020}. 

\begin{figure*}[]
    \centering
    \vspace{2mm}
\includegraphics[width=\linewidth]{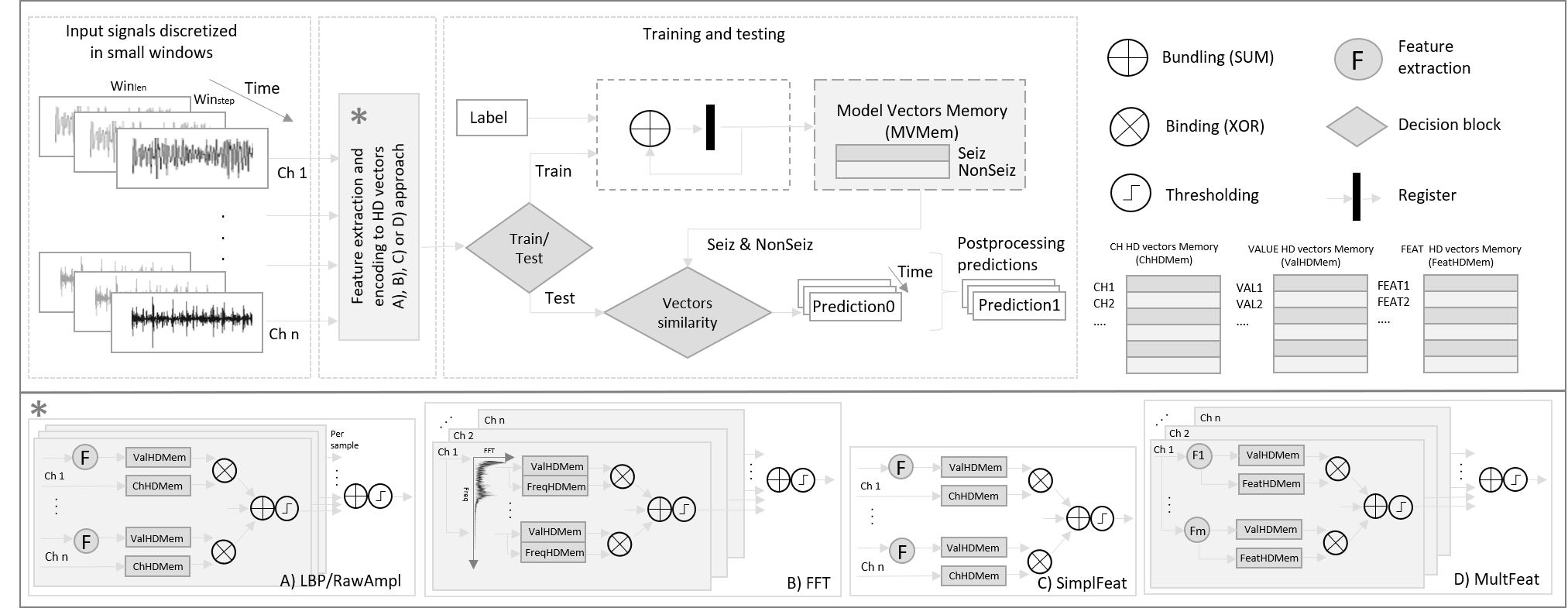}
    \caption{\small{Schematic of the HD computation workflow}} 
    \label{fig:workflowSchematic}
    \vspace{-8mm}
\end{figure*}

The first work that applied HD computing to epileptic seizure detection~\cite{burrello_one-shot_2018} relied on mapping local binary patterns (LBPs) to HD vectors. LBPs map a sequence of data samples onto a small binary array, relying solely on whether the data amplitude increases or decreases. 
The authors encode LBPs into HD vectors and test the approach on IEEG data from 16 patients from the Inselspital Bern epilepsy surgery program. They focused on testing one-shot learning, or learning from as few seizure instances as possible. They showed that for a majority of patients, the algorithm learned from one to two seizures and achieved perfect specificity and sensitivity. 

In another recent paper, HD computing was applied on EEG data, which is more viable for continuous long-term monitoring~\cite{asgarinejad_detection_2020}. The authors also compared HD computing with different standard state-of-the-art ML approaches (KNN, SVM, regression, random forests, and CNN). For KNN, SVM, regression, and random forests they used 54 different features from~\cite{zanetti_robust_2020} and~\cite{sopic_e-glass:_2018},
while for the HD computing approach raw amplitude values were encoded into HD vectors. More precisely, they normalized data and mapped amplitudes of all samples to the corresponding HD vector. The authors used the CHBMIT database from the Children's Hospital of Boston and MIT~\cite{shoeb_application_2009, goldberger_ary_l_physiobank_2000} 
and reported that the HD approach surpassed the performance of all other approaches.

Both aforementioned papers present promising results for the application of HD computing for epileptic seizure detection, but results cannot be directly compared. Namely, data preparation (filtering, train/test split), segmentation (size and step of discretized windows), mapping to HD vectors (how channel and time information is encoded) as well as performance metrics (sensitivity, recall, specificity, number of false positives) are different in both papers. If we extend the focus to papers applying HD computing for different problems, the diversity becomes bigger. For example, some works split the data into independent discretized windows and report a performance based on independent window labels, while others observe windows as a time sequence and also take seizure dynamics into account by performing additional post-processing steps to smooth the labels. Thus, performance measures can be based on the level of seizure episodes or considering the whole seizure duration. 
This is an ongoing topic of discussion, addressed in recent papers~\cite{ziyabari_objective_2019, shah_validation_2020}, specifically applied to epilepsy detection. 

Overall, unfortunately, it is often unclear in analysis of previous works why certain approaches have been used in specific applications and how other approaches would perform on the same task. This makes comparison between works very challenging. Thus, the main motivation of this work is to perform a systematic assessment of the HD computing framework for detection of epileptic seizures. In particular, we compare different feature encoding strategies on HD vectors by evaluating them in a comparable way using common workflow (i.e., with the same preprocessing setup and identical performance measures). We perform the analysis on two different datasets to assess the generalizability of our conclusions.

\section{HD framework and workflow}
\label{sec:OverallWorkflow}

\subsection{Training and testing workflow} 
\label{Subsec:TrainTestWorkflow}
Our complete analysis workflow is presented in Fig.~\ref{fig:workflowSchematic}. Data is discretized into windows of duration $W_{len}$ that are moved in steps of $W_{step}$, i.e., for every $W_{len}$ of data, features are calculated and encoded into an HD vector representing that discrete window of data. 
During the training step, windows belonging to one class are bundled into one HD model vector for each class. Bundling is achieved by performing bit-wise summation (SUM) over the HD vectors and rounding in the end. More specifically, if a majority of the vectors on that bit position have a value 1, the final value is 1, otherwise 0. 

For epileptic seizure detection, specifically, the previous approach leads in the end to two HD vectors: one for the ictal and one for the interictal class. Ictal relates to the part of the data where seizure was present, while interictal corresponds to the baseline EEG data distant from seizure episodes. Around ictal data, often pre-ictal and post-ictal phases are defined, but here we focused only on clear seizure (ictal) and non-seizure (interictal) classification. 

Finally, in the testing phase, the HD vector representing the current discrete window is compared with both model vectors, and the label of the more similar class is chosen. We use the Hamming distance between two vectors to measure similarity, as it is the most common metric for binary HD vectors.

\subsection{Feature encoding workflow}  
\label{Subsec:FeatEncodWorkflow}
First, HD vectors that represent different features $HDV_{Feat}$, feature values $HDV_{Val}$, and channels $HDV_{Ch}$ are created at the beginning of training in the form of a memory map, enabling simple one-to-one mapping between values and corresponding HD vectors. This means that if we have 10 features, 10 HD vectors are initialized, one representing each feature. Similarly, for channel vectors, one vector is initialized for each channel. For feature values, values are normalized and discretized to a certain number of levels, so that one vector is assigned to each normalized feature value. 

In this work, we use an approach where all $HDV_{Feat}$ and $HDV_{Ch}$ HD vectors are independently and randomly generated at the beginning of the training. On the other hand, $HDV_{Val}$ vectors were initialized in a way where first the vector was randomly initialized, but then every subsequent vector representing the next possible value was created from the previous one by permuting consecutive blocks of $d$ bits. The number of bits $d$ depends on the number of possible values (and corresponding $HDV_{Val}$ vectors) that are needed. This approach ensures that vectors representing numbers that have closer values are also more similar.

Next, feature encoding is performed by first calculating the feature(s) of interest for a given discrete window of size $W_{len}$. Then, for each feature value, an HD feature vector $HDV_{Feat}$ representing that feature, an HD value vector $HDV_{Val}$ representing the value itself, and the corresponding channel vector $HDV_{Ch}$ are selected. 

Since vectors are binary vectors, we bind vectors by using bit-wise XORs between each feature vector $HDV_{Feat}$ and the corresponding value of that feature vector $HDV_{Val}$ or channel vector $HDV_{Ch}$. All bound vectors are then bundled up (bit-wise summed (SUM) and rounded), resulting in an HD vector representing each discrete window.
Depending on the feature approach used, binding and bundling of data is slightly different, as described in Sec.~\ref{sec:AllApproaches}, as well as illustrated in Fig.~\ref{fig:workflowSchematic} (sub-windows A, B, C and D).

\subsection{Label post-processing}
\label{Subsec:LabelPostprocessing}
As data is discretized into windows $W_{len}$, but treated as time sequences, after predicting labels for each step $W_{step}$, we utilize time information to post-process (smooth) the predicted labels. For example, it is not realistic for seizures to last only a few samples, and two seizures closer than a few seconds are usually considered as the same seizure. Thus, we go through the predicted labels with a moving average window of a certain size $SW_{len}$ (5s) and perform majority voting. 
Further, if two seizures are closer than $DS_{min}$ (30s) we merge them (i.e., we flip all predictions between them from 0 to 1). Due to the small $W_{len}$ and even smaller $W_{step}$, granularity is much smaller than the dynamics of the seizures, so these steps lead to more realistic and robust seizure predictions. 

\subsection{Performance Evaluation}
\label{Subsec:PerfEvaluation}
In order to capture as much information as possible about the model performance, as proposed in ~\cite{ziyabari_objective_2019, shah_validation_2020}, we measure performance on two levels: 1) episode level, and 2) seizure duration level. 
Episode level detects seizure episodes, irrespective of whether the whole duration of a seizure was detected. On the other side, seizure duration level captures the seizure duration. 
For both, levels we measure sensitivity (true positive rate or $TPR$, calculated as $TP/(TP+FN)$), precision (positive predictive value or $PPV$, calculated as $TP/(TP+FP)$) and F1score ($2*TPR*PPV/(TPR+PPV)$). 

Metrics on these two levels give us better insight into the operation of the proposed systems. Furthermore, the performance measure often depends on the intended application and play a big role in the acceptance of the proposed technology. 

\subsection{Computational and Memory Complexity Evaluation}
\label{Subsec:CompMemEvaluation}

In order to characterize all the relevant aspects that could determine the feasibility of future implementations on embedded systems, we also study the memory requirements and computational complexity of the proposed feature approaches. 

Memory requirements are assessed by calculating the memory needed to store all model vectors for each particular approach (e.g., memory for $HDV_{Val}$ vectors, $HDV_{Feat}$ vectors, $HDV_{Ch}$ vectors, etc.). We express it relative to the dimensionality of the HD vectors $D$, i.e., actual memory size is calculated by multiplying by $D$ (number of values in each vector) and its type size (bit/integer/float). On the other hand, the computational complexity is estimated by calculating the number of SUM and XOR operations and measuring the time required for feature calculation and the time for encoding to an HD vector. It is measured per one discrete data window $W_{len}$, as all analysis and predictions are window-based. Time is assessed as the average time over 1000 discretized windows for both datasets and all feature approaches.

\section{HD Computing Approaches for Epileptic Seizure Detection}
\label{sec:AllApproaches}

In this section, we detail the different signal features that may be exploited for epileptic seizure detection, and how they are encoded into HD vectors. The training and testing procedures are the same for all approaches and follow the steps already explained in Sec.~\ref{Subsec:TrainTestWorkflow}.

\subsection{Local Binary Patterns}
\label{Subsec:LBP}
The \textit{LBP} approach maps a sequence of raw data into arrays of 0's and 1's. 
Namely, if the signal between two samples is increasing, the bit corresponding to that position has a value of 1, otherwise 0. This procedure maps the changes in the signal trend but not the values of the signal itself. The distribution of LBP codes within ictal and interictal states are different and pose an interesting feature to use: during the interictal states, LBP codes are almost evenly distributed over all possible codes, while in ictal windows certain codes are more dominant~\cite{burrello_one-shot_2018}. Each data window is represented by encoding LBP features into an HD vector. First, we calculate LBP binary patterns for each sample and map these binary vectors to the corresponding $HDV_{Val}$ HD vectors. We bind (XOR) $HDV_{Val}$ vectors with the corresponding $HDV_{Ch}$ channel vectors, yielding $HDV_{ValCh}$ vectors. $HDV_{ValCh}$  vectors are then bundled (summed and rounded) all together. This is repeated for all samples in each $W_{len}$ of data and bundled to represent that window as illustrated on Fig.~\ref{fig:workflowSchematic}(A). The length of the LBP patterns can be chosen arbitrarily, but here we use 6-bit patterns, as proposed in~\cite{burrello_one-shot_2018}. It was demonstrated in~\cite{burrello_hyperdimensional_2020} that this value leads to the best performance, and this also enhances comparability between papers.

\subsection{Raw Signal Amplitude}
\label{Subsec:RawAmpl}
The raw signal amplitude (\textit{RawAmpl}) approach is similar to the one of LBP because it does not rely on extracting any complex features from the data, but solely encodes normalized raw signal amplitudes into HD vectors. Namely, here we use a simpler version of the encoding from~\cite{asgarinejad_detection_2020} to make it more comparable to other approaches. First, we normalize the raw amplitude on a sample-by-sample basis. Then, we map the values to the corresponding $HDV_{Val}$ vectors and bind (XOR) together with $HDV_{Ch}$ HD channel vectors, yielding $HDV_{ValCh}$ vectors. This is repeated for each sample in a window. Finally, similar to Sec.~\ref{Subsec:LBP}, the HD vector representing the current window is computed by bundling (summing and rounding) all bound $HDV_{ValCh}$ vectors across time samples of that window.

\subsection{Frequency Spectrum}
\label{Subsec:FFT}
Since frequency spectrum components are common features used for epileptic seizure detection, we test whether encoding the FFT spectrum into HD vectors can be useful for classification. The \textit{FFT} approach we use is similar to the one in~\cite{imani_voicehd_2017}, i.e., applied on a voice recognition task. Here, as illustrated in Fig.~\ref{fig:workflowSchematic}(B) we calculate the FFT spectrum for each $W_{len}$. Then we bind (XOR) together normalized FFT values $HDV_{Val}$ with corresponding frequency HD vectors $HDV_{Freq}$, yielding $HDV_{ValFreq}$ vectors. This is repeated so that $Val-Freq$ vectors for all channels are bundled (summed and rounded) together to get an HD vector representing that data window.  

\subsection{Single Features}
\label{Subsec:SingleFeat}
Inspired by the simplicity of features from previous approaches, we also evaluate if single features calculated on $W_{len}$, and encoded to HD vectors can achieve similar performance as \textit{RawAmpl} and \textit{LBP}. We test three commonly used features in state-of-the-art ML algorithms for epilepsy detection:
\begin{itemize}
    \item \textit{Mean amplitude} is defined as the mean amplitude of the normalized signal in window $W_{len}$.
    
    \item \textit{Entropy} is defined as the normalized spectral entropy value in window $W_{len}$, and is defined as: 
    \begin{equation}
    \label{eqn:entropy}
        H(x,sf)=-\sum_{f=0}^{fs/2} PSD(f)*log_{2}[PSD(f)]
    \end{equation}
    where $PSD$ represents the normalized Power Spectral Density of the data, and $fs$ the sampling frequency.

    \item \textit{Continuous Wavelet Transform (\textit{CWT})} is defined as the area of the frequency spectrum with the highest energy. More precisely, we calculate the CWT decomposition of signal window $W_{len}$ with 20 levels in the range [0.25,15] Hz. The index of the frequency band with the highest energy is chosen as a feature.
    
\end{itemize}
    
For each feature, the feature value is mapped to HD vector $HDV_{Val}$ and bound (XOR) with the corresponding channel vector $HDV_{Ch}$. Finally, to get the HD vector representing a $W_{len}$ window of data, we bundle (sum and round) $HDV_{ValCh}$ vectors of all channels together, as shown in Fig.~\ref{fig:workflowSchematic}(C).

\subsection{Combining Multiple Features}
\label{Subsec:MultipleFeat}
Inspired by machine learning approaches where more features lead to performance improvements \cite{zanetti_robust_2020}, we combine multiple features in a single model. We test this in two steps: 
\begin{itemize}
\item \textit{3Feat}: utilizing the three above-mentioned features(mean amplitude value, entropy, and highest energy frequency band), and
\item \textit{45Feat}: using the 45 features proposed by a standard ML approach to seizure detection~\cite{zanetti_robust_2020}. 
\end{itemize}
These features contain 37 different entropy features, including sample, permutation, Renyi, Shannon, and Tsallis entropies, as well as 8 features from the frequency domain. 
For each signal window, we compute the power spectral density and extract the relative power in the five common brain wave frequency bands; delta: [0.5-4] Hz, theta: [4-8] Hz, alpha: [8-12] Hz, beta: [12-30] Hz, gamma: [30-45] Hz, and a low frequency component ([0-0.5] Hz). These features are commonly held to be medically relevant for detecting seizures~\cite{teplan_fundamental_2002}.

For each feature, its value $HDV_{Val}$ and its index vector $HDV_{Feat}$ are bound (XOR), to get $HDV_{ValFeat}$ vectors. Finally, to get a final HD vector representing each $W_{len}$ , we bundle (sum and round) $HDV_{ValFeat}$ vectors of all features and channels, as shown in Fig.~\ref{fig:workflowSchematic}(D).

\section{Experimental Setup}
\label{Sec: Experimental_setup}

\subsection{Databases}
\label{Subsec:databases}

We assess the studied approaches on two publicly available databases: one for EEG data~\cite{shoeb_application_2009, goldberger_ary_l_physiobank_2000} and one for IEEG data~\cite{burrello_one-shot_2018}. We are interested to see if the trends for different performance measures will be the same for both databases.

The CHB-MIT database is an EEG dataset consisting of 24 subjects with medically resistant seizures ranging in age from 1.5 to 22 years~\cite{shoeb_application_2009, goldberger_ary_l_physiobank_2000}, and is collected by the Children's Hospital of Boston and MIT. We use the 18 channels that are common to all patients. Overall, the dataset contains in total 183 seizures, with an average of 7.6 $\pm$ 5.8 seizures per subject. 

For IEEG, we use a database of 16 patients from the Inselspital Bern epilepsy surgery program, which was also used in~\cite{burrello_one-shot_2018}. The database contains in total 100 seizures, with an average of 6.3 $\pm$3.8 seizures per subject. Each recording consists of a 3-minute interictal segment, an ictal segment (ranging from 10 to 1002 seconds), and finally, a 3-minute interictal segment. The number of channels was variable per patient due to differences in implanted IEEG electrodes.  

\subsection{Dataset Preparation}
\label{Subsec:datasetPreparation}

Several steps are required to make the databases comparable and applicable for epileptic seizure classification. First, both databases are (re)sampled to 256Hz. Further, we create balanced files for each seizure, where, in each file, there is an equal amount of interictal (non-seizure) and ictal (seizure) data. More specifically, from the original database files, for each file containing ictal data, we keep all ictal data. Before the onset of the seizure, we keep the same length of randomly selected interictal data. We exclude interictal data that is within 1 minute of seizure onset and up to 15 minutes after a seizure, as this data might contain ictal patterns. Balancing the data allows us to focus on assessing the separability between classes and prevents issues related to the prior distributions among classes. 

\subsection{Validation}
\label{Subsec:evaluation}
Given the subject-specific nature of seizure dynamics, we evaluate the performance on a personalized level. As already mentioned, the data for each subject is divided into files, where each file contains one seizure and the same amount of non-seizure samples. This enables us to use a leave-one-seizure out approach, where the HD model is trained on all but one seizure/file, while at the same time ensuring a balanced dataset for both training and testing. For example, for a subject with $N_{seiz}$ files (each containing one seizure), we perform $N_{seiz}$ leave-one-out cross-validations and measure the final performance for that subject as the average of all cross-validation iterations. 

All the code and data required to reproduce the presented results are available as open-source \footnote{https://c4science.ch/source/HDforEpilepsyPublic/}.

\section{Experimental Results}
 
\subsection{Prediction performance }
\label{Subsec:predictionPerformance2}

\begin{figure}[]
    \vspace{2mm}
    \centering
\includegraphics[width=\linewidth]{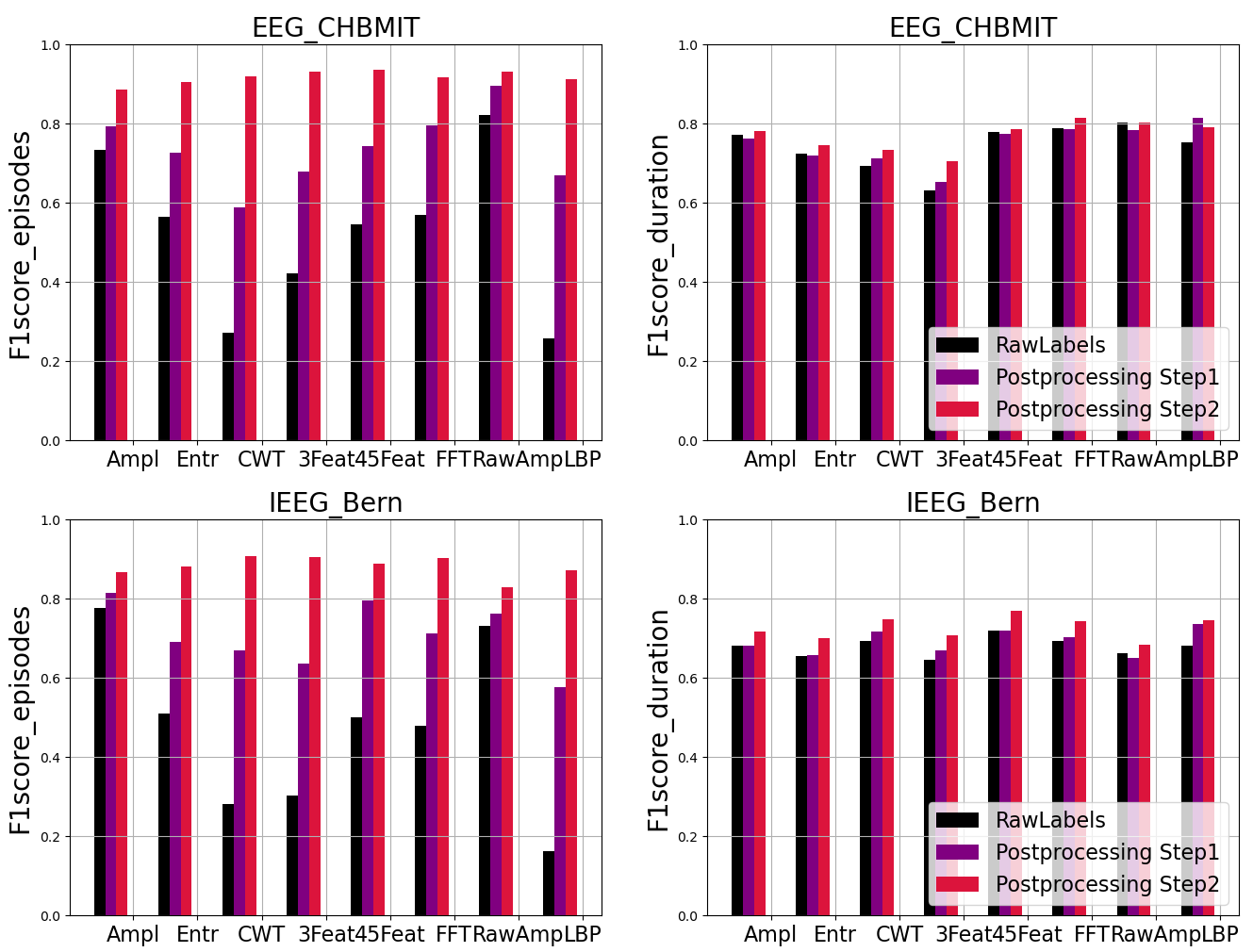}
    \caption{\small{F1 score performances for different feature approaches, both for episodes and duration level and two databases.}} 
    \label{fig:F1scorePerformance}
    \vspace{-2mm}
\end{figure}

Fig.~\ref{fig:F1scorePerformance} shows F1 score of epileptic seizure detection before and after two steps of post-processing for both databases. 
Performing moving average with voting in the first step improves performance at the episode level (up to 41.2\% for EEG and 41.5\% for IEEG). However, adding another step to merge close seizures improves the performance even more (up to 65.5\% for EEG and 70.9\% for IEEG). On the other hand, post-processing steps provide a minor improvement on the seizure duration level (up to 4.0\% for EEG and 6.4\% for IEEG after both steps). Papers in the literature use different post-processing approaches (often without clearly mentioned parameters); thus comparison among different works is difficult. 

We compare feature approaches primarily before label post-processing, separating it from the influence of post-processing. At the episode level, the best performance is achieved by approaches including amplitude information (\textit{RawAmpl} and \textit{Ampl}), while \textit{Entropy} and \textit{CWT} perform worse than \textit{Ampl}. Interestingly, combining the three features does not lead to a performance improvement over just individual features itself. Further, using 45 features \textit{45Feat} rather than three \textit{3Feat} leads to an improved performance, but not better than \textit{Ampl} alone. \textit{FFT} performs on the level of 45Feat, while \textit{LBP} performance is much lower. The range of F1 score performance before post-processing for different feature approaches is [25.8\%, 82.2\%] for EEG, and [16.3\%, 77.7\%] for IEEG. This is due to the large number of false positives that exist before post-processing. After post-processing, differences between approaches are significantly reduced, i.e., with range [88.6\%,  93.7\%] for EEG and [82.8\%, 90.7\%] for IEEG.

Performance on the level of duration is, on the other hand, less variable between approaches, with the F1 score ranging [63.0\%, 80.4\%] for EEG and [64.4\%, 72.0\%] for IEEG, before post-processing. Moreover, improvement with post-processing is smaller, i.e., performance is [70.4\%, 81.4\%] for EEG and [68.3\%, 76.9\%] for IEEG. 
The relative performance between different feature approaches have very similar trends on both databases, which makes results more generalizable in the scope of epileptic seizure detection.

Finally, when comparing the final F1 performance after two steps of post-processing,  
approaches differentiate in 5.1\% and 11.0\% for episodes and duration of EEG database, and similarly, 7.9\% and 8.6\%, for IEEG database. 
Here, we see that after post-processing, approaches differ less on the level of episodes, even though they differ more before post-processing. This means that the best choice of feature approach to use, depends not only on the amount of post-processing but also the performance objective (i.e., only seizure episode detection or also seizure duration detection).

\begin{table*}[t]
\vspace{3mm}
\caption{Memory and computational requirements}
\begin{center}
\begin{tabular}{|c|c|c|c|c|c|c|c|c|c|}
\hline
\textbf{Dataset}& \textbf{Performance type} &\textbf{Ampl} &\textbf{Entr}&\textbf{CWT}&\textbf{3Feat}&\textbf{45Feat}&\textbf{FFT}&\textbf{RawAmpl}&\textbf{LBP} \\
\hline
& \textbf{Memory (factor of D)} & 36 & 36 & 36& \textbf{23} & 65 &84 &36 & 80\\
& \textbf{Number of operations} &\textbf{32} & \textbf{32} & \textbf{32} & 112 & 1456 & 2064 & 33792 & 33792\\
\textbf{CHBMIT} &\textbf{Computation time [ms]} & \textbf{0.00093} &0.0061 & 0.3343 & 0.3769 & 0.3766 &0.0108 & 0.5788 &0.3883 \\
&\textbf{Ratio Feat vs HDvec} & 1.809 &15.618 &690.328 &33.544 &33.988 & 0.583 & 0.787 &0.587\\
\textbf{IEEGBern} &\textbf{Computation time [ms]} & \textbf{0.001} & 0.0064 & 0.3277 & 0.3555 & 0.3530 & 0.0109 & 0.5573 & 0.3787\\
&\textbf{Ratio Feat vs HDvec}  &1.657 & 14.690 & 617.180 & 34.788 & 31.525 & 0.574 & 0.777 & 0.583\\
\hline
\end{tabular}
\label{tab1:Memory&computPerf}
\end{center}
\vspace{-8mm}
\end{table*}

\begin{figure}[]
    \vspace{2mm}
    \centering
\includegraphics[width=\linewidth]{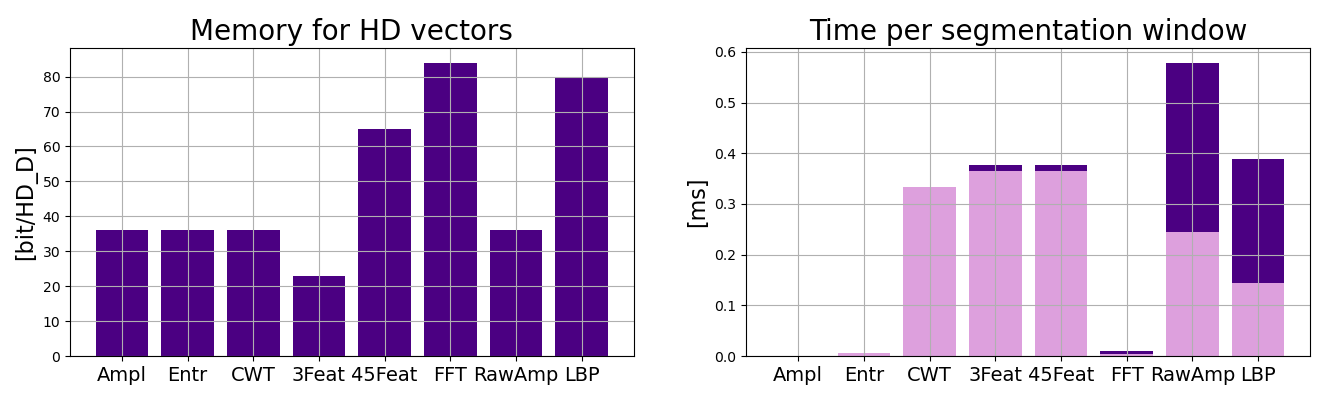}
    \caption{\small{Memory and computational complexity for different feature approaches. Right graph represents total time for processing one discrete window of data. The pink color represents the amount of time needed for calculating the feature values and the blue portion is the time required for operations on HD vectors.}} 
    \label{fig:MemAndComputRequirements}
    \vspace{-2mm}
\end{figure}

\subsection{Computational Complexity and Memory Requirements} \label{Subsec:CompAndMemRequirements}

The complexity and memory requirements results are shown in  Fig.~\ref{fig:MemAndComputRequirements} and Table~\ref{tab1:Memory&computPerf}. Memory requirements for storing HD vectors are different among approaches. They depend on the exact number of channels, features, segmentation levels for feature values, and sampling frequency. 

The memory requirements are expressed in Table~\ref{tab1:Memory&computPerf} as a factor of the dimensionality of HD vectors, $D$. To get actual memory in bits we need to multiply it with the number of bits $D$ that form base vectors. \textit{LBP}, \textit{FFT} and \textit{45Feat} approaches have biggest requirements; \textit{LBP} due to the large number of possible \textit{LBP} values, \textit{FFT} due to high frequency resolution and \textit{45Feat} due to the large number of features.   

In terms of computational complexity, variance is very high between different approaches. The most computationally expensive is \textit{RawAmpl}, followed by \textit{LBP}, and then \textit{45Feat}, \textit{3Feat} and \textit{CWT}, but the reasons for this complexity are different among approaches. In  Table~\ref{tab1:Memory&computPerf}, the ratio of time required for feature calculation vs. time to perform all operations on vectors, as well as the total time, is shown.
For example, \textit{RawAmpl} and \textit{LBP}, even though they are simple to calculate, are sample-based. Thus, due to their high sampling frequency, the number of operations in one discretized window is very high (33792 XOR+SUM operations vs. 32 for single feature approaches). 
This is also visible in Fig.~\ref{fig:MemAndComputRequirements}, as a big portion of the bars are blue, representing time needed for operations on vectors, i.e., 56\% for \textit{RawAmpl} and 63\% for \textit{LBP}. 
On the other hand, \textit{CWT}, \textit{3Feat} and \textit{45Feat} are more computationally consuming due to the complexity of the calculated features. As shown in  Fig.~\ref{fig:MemAndComputRequirements}, a majority portion of the corresponding bars (>97\%) are pink, representing feature value calculation. In contrast, \textit{Ampl}, \textit{Entropy}, and \textit{FFT} are very computationally efficient, both due to a lower number of operations on vectors and a  simpler feature calculation. 

Overall, the ratio between the best and worst-case memory requirements are $3.8\times$, $1056\times$ for number of operations, and $622\times$ (EEG) and $557\times$ (IEEG) for the computation time. Thus, differences between approaches are much smaller concerning memory than for computational complexity.

Finally, our results show that approaches with high performance (such as \textit{RawAmpl} or \textit{45Feat}) are not ideal for wearable applications due to high memory or computational requirements. At the same time, similar but simpler approaches (e.g., single feature for \textit{Ampl}) can maintain high precision and require significantly less memory and computational resources. Thus, for HD computing, similar to standard ML, feature selection is an extremely important step to properly optimize the performance and hardware implementation feasibility.

\section{Conclusion}

In this work, we have performed a systematic assessment of the HD computing framework for the detection of epileptic seizures. In particular, we compared different feature encoding strategies on HD vectors. Also, as HD computing is a novel approach and there is a lack of guidelines on strategies, parameters and calculation steps, we have systematically analysed the workflow in a practical application, with emphasis on feature encoding and multi-channel fusion. More precisely, we tested two known feature methodologies (\textit{LBP} and \textit{RawAmpl}), as well as several novel approaches for HD computing on epileptic seizure detection (Single Feature - \textit{Ampl}, \textit{Entr}, \textit{CWT}; Multiple Feature - \textit{3Feat} and \textit{45Feat}, and \textit{FFT}).

In our experiments we compared three main aspects: 1) detection performance, 2) memory requirements and 3) computational complexity. Results show a difference in F1 score between approaches (up to 61.4\% for episodes detection and up to 17.4\% for seizure duration), but also that the highest performance approaches are not ideal for wearable applications due to their high memory or computational requirements. Nonetheless, after post-processing steps for smoothing prediction labels to adjust predictions to the dynamics of epileptic seizures, performance is improved and differences are significantly reduced between approaches (up to 7.9\% for episodes detection and up to 11.0\% for seizure duration).  In terms of memory and computational requirements, differences between approaches are much smaller concerning memory ($3.8\times$) than for computational complexity ($622\times$). Thus, for a wearable implementation, feature selection and decision based on several aspects are necessary. 
All the analysis results were performed and confirmed on two different datasets, which improves the generalizability of the results.

\printbibliography

\end{document}